# Manufacturing processes hardware and software development to implement innovative technologies of aircraft manufacturing facilities management

**S.E. Pyatovsky**

*PhD in Economic sciences, Associate Professor of the Department of «Management of high-tech enterprises», Moscow Aviation Institute (National Research University)*

*(e-mail:vgsep@ya.ru)*

**A.N. Serdyuchenko**

*Expert (Category I), Moscow Aviation Institute (National Research University)*

*(e-mail: anton415@gmail.com)*

Abstract. *The paper presents approaches to modern aircraft manufacturing facilities' competitiveness growth based on innovative technologies of management decisions implementation. The paper establishes a connection between the aircraft industry and the development of the national economy, along with the dependence of the Russian economy on international manufacturers of civil airplanes. Comparative statistics of civil aircrafts produced by Russian and international manufacturers are given. Comparative analysis of military and civil aviation is done. It is shown that project implementation based on Open Source and OLAP-technologies at a high-technology enterprise is a pre-requisite for the competitive growth of aircraft manufacturing facilities.*

Keywords: *innovative technologies, management decision making support system, high-tech enterprise competitiveness, OLAP, Open Source projects, manufacturing processes hardware and software*

High-end industry forms the basis of highly developed countries' economies. However, for the time being, the civil airplanes manufacturing industry is declining [1,2,3]. To reverse the trend, Russian aircraft manufacturing companies' competitiveness in the international market needs to be enhanced. Management decisions quality improvement based on modern technologies of manufacturing facility operation analysis, like OLAP-technologies, is one of the ways to enhance competitiveness. The purpose of the article is to describe the development of the system for improving the quality of managerial decisions based on innovative technologies that analyze the activities of the civil aviation enterprise.

**1. Aircraft manufacture impact on the national economy**

Aviation is one of the major means of modern transport, with a dominant role in numerous branches of industry, which in the long run facilitates national income. According to NPO's Air-Transport Action Group (ATAG) data, if aviation were a state, it would be the nineteenth state in the world in the terms of GDP, making about $540 billion per annum. This points out an essential role of the national aircraft manufacturing industry in the national economy [4]. This relation is recognized by scientific researchers (F. Bourguignon, P.E. Darpeix) [5] and others, who draw the conclusions that the aircraft manufacturing industry's status depends on the level of economic activity along with the population's well-being level, that is, aircraft manufacturing facilities as high-end enterprises are indicators of the states' national development.

**2. Comparative study of the Russian and international civil aircraft manufacturing industries**

United Aircraft Corporation (UAC), the biggest association of aircraft manufacturing companies in the country, is a key player in the aircraft manufacturing industry of Russia.

Table 1 presents a comparative history of civil aircrafts supplied by UAC and leading transnational corporation.

Table 1

**Civil aircrafts supplied [6,7,8] by UAC, Airbus and Boeing**

| Year | 2014 | 2015 | 2016 | 2017 |
|---|---|---|---|---|
| **UAC** | 37 | 29 | 37 | 0 |
| **Airbus** (% UAC to Airbus) | 629 (5.9) | 635 (4.6) | 688 (5.4) | 306 (0) |
| **Boeing** (% UAC to Boeing) | 723 (5.1) | 762 (3.8) | 748 (4.9) | 352 (0) |

As Table 1 shows, UAC is significantly behind international leaders. As of the beginning of 2017, aircraft manufacturing facilities' condition can be characterized as in a recession. For the first three months of 2017, there were not any sales proceeds (sales results) from helicopters, aircrafts, or other airborne devices [8].

The current situation requires from aircraft manufacturing facilities a fresh approach to innovative project arrangement in management decision-making gathering and support system (MDMGSS). The importance of these innovative projects was considered in academic papers (J. Daily, J. Peterson) [9], (V.A. Komarov, S.A. Piyavskiy) [10], as well as by leading aircraft companies' management, like Ted Colbert (IT and Data Analysis Director, Boeing) [11], David Kasik (Technical Expert in visual simulation and interactive technologies, Boeing) [12], and others.

Russian aircraft manufacturers' need to improve competitiveness is acknowledged by research papers on the necessity of IT implementation at high-tech manufacturing facilities [13], as well as by the developed national program "Aircraft manufacturing industry development" [14]. The program is focused on "highly competitive aircraft industry creation and its position at international market consolidation as 3$^{rd}$ player in terms of aircrafts (airborne vehicles) output rate" [14].

**3. National economy's dependence on international manufacturers of civil aircrafts**

The recession state of Russian civil aviation causes strong dependence of the national economy on international aircraft manufacturers. For example, the Aeroflot company, which lead in terms of passenger traffic in 2017 [15], has an aircraft fleet of 198 airplanes, the majority of which are Airbus and Boeing aircrafts [16] (Table 2).

Table 2

**Aeroflot airline company aircraft fleet**

| Airplane model | B777 | B737 | A330 | A321 | A320 | SSJ-100 |
|---|---|---|---|---|---|---|
| Aircrafts number | 16 | 26 | 22 | 36 | 68 | 30 |

It appears from Table 2 that intense cash flows, which could be invested in the Russian economy, go to overseas aircraft companies. Strong dependence on transnational manufacturers is evidenced by the fact that Russia "being on the needle of foreign aircraft industry spends and pays abroad 470 billion RUB annually."

**4. Military and civil aviation benchmarking**

Due to the heavy expenses of the developed countries on the defense industry [17], it is necessary to examine defense investments as an incentive for civil production development (Tables 3 and 4).

Table 3

**Countries in the terms of military expenditures [18], air forces (AF) military staff number and military aircraft number [19]**

| Countries | | Military expenditure in 2015[*], $ billion | AF military staff number | Military aircraft number |
|---|---|---|---|---|
| USA | | 596.0 | 2 363 675 | 13 762 |
| Europe | France | 50.9 | 387 635 | 1 305 |
| | Germany | 39.4 | 210 000 | 698 |
| | UK | 55.5 | 232 675 | 856 |
| | Spain | 14.1 | 174 700 | 533 |
| | Total | 159.9 | 1 005 010 | 3 392 |
| Russia | | 66.4 | 3 371 027 | 3 794 |

[*] *Table presents total military expenditures, not Air Force, due to restricted data.*

Table 4

**Civil aircraft manufacturing facilities' turnover, employment rate and manufactured aircrafts number**

| Aircraft manufacturing | Civil aircraft manufacturing facilities, $ billion | Employment rate on civil aircraft manufacturing facility | Number of civil aircrafts manufactured in 2016 |
|---|---|---|---|
| Boeing | 96 | 165 500 | 748 |
| Airbus | 36 | 133 000 | 688 |
| UAC | 7 | 100 000 | 37 |

Tables 3 and 4 confirm that overlarge defense branch maintenance [20] in economic slack circumstances is quite ineffective budget allocation [21]. Russia exceeds Europe in terms of numbers of AF staff members and military aircrafts; however, Airbus exceeds UAC in terms of sales turnover, number of employees at the manufacturing facility, and number of manufactured civil aircrafts. The state "spends [for defense] fair amount. These expenditures [in 2016] reached 3%. Efficiently functioning economy is of major importance for the state. Thus, army must be of moderate size, however up-to-date and efficient." [22]

In such a way, for the sake of investment effective allocation, investments in military infrastructure shall satisfy efficiency, flexibility, and processability criteria.

**5. High-tech aircraft manufacturing facility cost efficiency improvement**

Increasing the competitiveness of an enterprise in the modern market is associated with an increase in the quality of management decision-making. Company management quality is established at the legislative level, including international systems of management quality having Russian substitute detailed defined in state standards system.

GOST R ISO "Quality management systems" [22] contains a number of ways to increase a manufacturer's performance factors, including "Decisions making based on evidences" (p. 7), which contains the requirements for data analysis quality, such as drafting, substantiation, key benefits, and possible actions.

For a qualitative analysis of information, an aviation enterprise must use innovative technologies. Innovative technologies provide the best solutions that meet the requirements of the highly competitive market [23]. A high-tech enterprise operating in the international market does sale and operations analysis to maximize management efficiency, cut expenditures, and gain market share. The more data the enterprise is able to collect of the industry branch and economy in its entirety having impact on the company operating results, the more likely that data will increase the quality of management decisions and the competitiveness of the enterprise.

High-tech enterprises are developing the automated economic information systems (AEIS). The developed AEIS framework involves data acquisition required and sufficient to generate prices, reliable and immediate information of the company activities, that means management system non-redundant and sufficient information space creation. A manufacturing facility able to convert data into knowledge and create a knowledge database (KDB) in its information contour in a precise and timely manner will be better prepared to make successful management decisions and increase competitiveness.

To reach maximum efficiency, an AEIS system shall employ highly efficient yet cheap tools of data analysis. Such tools include OLAP-technology, a powerful and convenient tool for data analysis [24].

### 6. Open Source projects

Open Source projects, representing Open Source software (soft) available for review, examination, and alteration, are used to provide minimum expanses for project automation at large-scale high-tech manufacturing facilities. Using Open Source applications, aircraft manufacturing industry facilities have the opportunity to use powerful tools without investment in expensive software licenses. In particular, solutions development based on a Mondrian server such as Open Source OLAP [25] has proved its efficiency at aircraft manufacturing facilities. Mondrian is developed and maintained by Pentaho Corporation (Orlando, USA). Mondrian is released under the EPL-license, which is free software, whereby any developer has the opportunity to participate in the creation of this software. A free license allows using the software regardless of the conditions of the sanction economy.

### 7. Decision Support System (DSS) arrangement based on OLAP-technologies and Mondrian solutions

Analysis and management decision-making is accomplished based on the reports with variables in the following spreadsheet tables. Such an approach makes it easy to display resulting data as well as to work with 2D (standard spreadsheets) and 3D (books with same report tabs on various variables) measurements. However, an aircraft manufacturing facility manages its activity processing set of variables. Such complicated activities analysis pushes the boundaries of linked dBase tables and is not suitable to open and display new data as a good many resources are required to create a new report.

As opposed to relational databases (DB), OLAP-technologies do not store separate transaction records in two-dimensional format by tuples; however, they use DB multidimensional structures to store consolidated data arrays [26]. OLAP-technologies provide fast access to aggregate multidimensional information. OLAP-technology allows one to create fast clusters and basic data packages computation in view of DSS establishment. In such a way, OLAP-technologies provided the base for planning, analysis, and reports, allowing analytics to create efficient business models of high-tech manufacturer.

Being by its nature a relational mechanism, Open Source OLAP (ROLAP), Mondrian provides for data storage in relational database converting MDX queries in SQL-queries for data base management systems (DBMS). DWH data stores industry has shown that data organization in star topology allows for fast analysis of a large bulk of data. This is explained by the fact that relationships among the data is simplified and the number of links required for data aggregation comes to minimum.

Data in DWH are filled using the ETL (extract, transformation, load) process, initially withdrawn from source systems featuring relational databases and big data systems, like Hadoop, NoSQL, and MongoDB. Data are converted to enter the DWH system. The last one can contain stages like data cleansing and modification. The final data is loaded into DWH, where they are used by Mondrian software package.

One needs to create a complicated SQL-query to receive necessary data from standard schema. On the other hand, star schema has a structure easier to comprehend, simplifying SQL-query making. After Mondrian complex use, a decision maker gets an opportunity to perform data analysis without the needing to have knowledge of SQL.

**Conclusion**

Relations between the aircraft industry and national economy are reviewed and shown. The method for increasing the competitiveness of a high-technology aviation enterprise by developing a system for supporting management decision-making is presented. The developed system management decisions quality is done based on:

1. OLAP-technologies with Open Source projects and Mondrian solutions implementation at high-tech manufacturing facilities;
2. Data analysis quality improvement for management decisions support.